\documentstyle[preprint,revtex]{aps}
\begin{document}
\draft
\preprint{IPT-EPFL October 10,1994}
\begin{title}
Exactly Solvable Kondo Lattice Model
\end{title}
\author{D. F. Wang and C. Gruber}
\begin{instit}
Institut de Physique Th\'eorique\\
\'Ecole Polytechnique F\'ed\'erale de Lausanne\\
PHB-Ecublens, CH-1015 Lausanne-Switzerland.
\end{instit}
%%%%%%%%%%%%%%%%%%%%%%%%%%%%%%%%%%%%%%%%%%%%%%%%%%%%%%%%%%%%%%%%
\begin{abstract}
In this work, we solve exactly a Kondo lattice model in the
thermodynamic limit. The system consists of an electronic conduction
band described by unconstrained hopping
matrix elements between the lattice sites. The conducting
electrons interact with a localized impurity spin at each
lattice cell. We have found the exact thermodynamics,
the ground state energies of the system.
At $T=0$, we explicitly demonstrate that
the system exhibits a metal-insulator phase transition at half-filling.
In the limit of strong coupling between the impurity spin
and the conduction electrons, $J=+\infty$, we have solved the system
on a lattice of any size $L$. The ground states are the RVB-type
Jastrow product wavefunctions. Various correlation functions
can be computed for the impurity spins, and for
the singlets formed by electrons and impurities.
\end{abstract}

%%%%%%%%%%%%%%%%%%%%%%%%%%%%%%%%%%%%%%%%%%%%%%%%%%%%%%%%%%%%%%%%
\narrowtext
The heavy-fermion systems have been of great theoretical interests
in recent years\cite{lee,flude,coleman1,nick,coleman2,rice}.
A basic model for a heavy fermion system is
an $S=1/2$ Kondo lattice with a conducting band interacting with
a localized impurity spin at each unit cell. Quite recently,
a possibility has been discussed that heavy-fermion superconductors
involve odd-frequency triplet pairing, through a Majorana
representation for the local moments to
avoid a Gutzwiller projection\cite{coleman2}.
In spite of several numerical study and approximate
approaches to the Kondo lattice model, it is too difficult to
find the exact solutions for the system.

In one dimension, the Kondo lattice model has been studied through
various numerical and approximate methods. The results
indicate the existence of a finite spin gap and a finite charge gap
simultaneously at half filling\cite{fye,ts,wang,yu,tsvelik}.
The numerical renormalization group method can even
deal with a finite lattice up to 24 sites\cite{yu},
confirming further the charge gap at half-filling in
the thermodynamic limit. It also shows that
the charge gap is larger than the spin gap at half-filling\cite{ts,wang}.
The finite charge gap shows that the system described
by the Kondo lattice model at half-filling is an insulator.
However, due to the complexity, it has been
impossible to prove the existence of the metal-insulator phase
transition at the half-filling through any exact approach
without employing an approximate scheme or numerical methods.

The single impurity Kondo model can be reduced to a one
dimensional problem, as only the S-wave scattering off
the impurity spin is important.
The single impurity Hamiltonian, which represents a local Fermi liquid,
is exactly solvable with the Bethe
ansatz if the conduction band is described by continuum electrons with
linear spectrum at the Fermi surface\cite{andrei1,andrei2}.
The Bethe-ansatz solutions
explicitly demonstrate the dynamical
energy scale that enters the expression of
the thermodynamics of the system\cite{anderson,andrei1,andrei2}.
The screening of the impurity spin by conduction electrons,
the crossover from weak coupling fixed point to strong coupling
fixed point behavior can be explicitly illustrated with the Bethe-
ansatz solutions\cite{andrei1,andrei2}.

Despite the Bethe-ansatz solutions for the single impurity Kondo
model, no exact solution for more than one impurity spin is
available by far, either in continuum case or on a lattice.
In this work, we introduce a Kondo lattice model in which
the conduction electrons hop with unconstrained hopping
matrix elements. The local moment at each site interacting
with the conducting electrons is described by a spin 1/2.
We solve this Kondo lattice system exactly in the thermodynamic
limit at any temperature.
Although the hopping matrix of the conducting electrons
is far from the realistic case ( electrons hop between
the nearest neighboring sites in the tight binding picture),
our model provides the first
example where its exact solutions explicitly demonstrate
the metal-insulator phase transition at half-filling, as
the interaction between the local impurity moments and the
conduction band is turned on.

The Hamiltonian of the system is given by:
\begin{equation}
H=\sum_{1\le i\ne j \le L}\sum_{ \sigma=\uparrow,\downarrow}
t_{ij} c_{i\sigma}^\dagger c_{j\sigma} + J \sum_{i=1}^L
c_{i\alpha}^\dagger {\vec \sigma_{\alpha\beta} \over 2} c_{i\beta}
\cdot {\vec S_f(i)},
\end{equation}
where the hopping matrix element for the conduction electrons
is given $t_{ij}=-t$ for all $i\ne j$. $J$ is the coupling
constant between the local impurity moments and the conducting
electrons. We assume that $t>0$.
The electron creation and annihilation operators
have the usual anticommutation relations $\{c_{i\alpha},c_{j\beta}\}_+
=0, \{c_{i\alpha}^\dagger, c_{j\beta}\}_+ =\delta_{ij} \delta_{\alpha\beta}$.
The local moments are described by the spin 1/2 operators, that is,
$[S_f^x(k), S_f^y(k)]=iS_f^z(k)$ (plus two other commutation
relations obtained by the cyclic permutations of $x,y,z$),
with the relation $\vec S_f^2(k) =3/4$, for all the sites $k=1, 2, \cdots, L$.
Because of the special form of the conduction band, the dimensionality
of the lattice is irrelevant, and the system is basically one
dimensional. In the following, we always
discuss the thermodynamic limit where the lattice size
$L\rightarrow \infty$.

The Hamiltonian of the system has $SU(2)$ spin symmetry and
the $SU(2)$ isospin symmetry, if we represent the local impurity
moment through an $f$ fermion: $\vec S_f (k) = f_{k\alpha}^\dagger
{\vec \sigma_{\alpha\beta} \over 2 } f_{k\beta}$ at each site $k$
with the condition that $\sum_{\alpha} f_{k\alpha}^\dagger f_{k\alpha}=1$.
The isospin operators are given by $I^{+} =
\sum_i (-1)^i (c_{i\uparrow}^\dagger c_{i\downarrow}^\dagger
-f_{i\uparrow}^\dagger f_{i\downarrow}^\dagger),
I^{-}=(I^\dagger)^\dagger, I_z=\sum_{i\sigma}(c_{i\sigma}^\dagger c_{i\sigma}
-f_{i\sigma}^\dagger f_{i\sigma}-1)/2$\cite{jones}.
In the Fourier space, the conduction electron Hamiltonian takes
the following form
\begin{equation}
H_c=\sum_{1\le i\ne j \le L}
\sum_{\sigma=\uparrow,\downarrow} t_{ij}c_{i\sigma}^\dagger
c_{j\sigma} =\sum_{k}\sum_{\sigma=\uparrow,\downarrow}
\epsilon(k) c_{k\sigma}^\dagger c_{k\sigma} ,
\end{equation}
where the Fourier transform of the fermion operators are
$c_{k\sigma}^\dagger = {1\over L^{1/2}} \sum_{i=1}^L
e^{ikx_i} c_{i\sigma}^\dagger,
c_{k\sigma}={1\over L^{1/2}} \sum_{i=1}^L e^{-ikx_i} c_{i\sigma}$.
The wave vectors are given by $k=2n\pi/L$, with
$n=-(L-1)/2, \cdots, (L-1)/2$ .
The dispersion of the conduction band
is $\epsilon(k) = (-t'L) \delta_{k, 0} +t $. Here,
$t'=t$, however, we treat this as an independent parameter in the following.
The dispersion indicates that there are energy levels
$(-t'L+t)$ and $t$, and the plane waves with energy $t$ have
large degeneracies.
Because of this, to find the thermodynamics of this Kondo lattice
model, we shall employ the argument of van Dongen and Vollhardt\cite{dongen}
for the Hubbard model.
The impurity spin interaction
$H_{im}^J=J \sum_{i=1}^L
c_{i\alpha}^\dagger {\vec \sigma_{\alpha\beta} \over 2} c_{i\beta}
\cdot {\vec S_f(i)}$
can be treated as a perturbation, and
the thermodynamic grand potential may be expanded as a power series of the
coupling constant $J$. In the thermodynamic limit, we may see that
the grand thermodynamic potential comes from two independent parts,
with a correction of relative order $O(1/L)$ or $O(1)$,
which is ignorable
in the thermodynamic limit.

The partition function is given by
\begin{equation}
Z(t, J) = Tr[T \exp\{-\int_{0}^\beta d\tau (H(\tau)-\mu N(\tau))\} ],
\end{equation}
where $N$ the number of electrons, $\mu$ the
chemical potential,
and the grand thermodynamic potential is $\Omega = -\beta^{-1} \ln Z$.
The trace is over the Hilbert space for electrons and impurities.
Treating $H_{im}^J$ as a perturbation term, we write
$\Omega = \Omega_0 - \beta^{-1} W_{t'} (\beta, J)$, where $\Omega_0
=-\beta^{-1}
 \ln Tr exp({-\beta(H_c-\mu N)}) $. The contributions from all the connected
diagrams are given by
\begin{equation}
W_{t'}(\beta,J)=<T \exp[-\int_{0}^\beta d\tau H_{im}^J(\tau)]>_0^c.
\label{eq:key}
\end{equation}
Here, the notation of the expectation value $<T A(\tau_1) A(\tau_2)
\cdots A(\tau_n)>_0^c$ is defined
to be the sum of all the connected diagrams
of
\begin{equation}
<T A(\tau_1) \cdots A(\tau_n)>=
Tr T e^{-\int_0^\beta d\tau (H_c -\mu N)} A(\tau_1) A(\tau_2)
\cdots A(\tau_n)
/Tr e^{-\int_0^\beta d \tau (H_c - \mu N)}.
\end{equation}
The leading contributions to the quantity $W_{t'}(\beta,J)$ only come from
those connected diagrams consisting of the bare electron propagators
with momenta $k\ne 0$ scattering off the impurity spins.
The electron propagator
with momentum $k=0$ has $t'$-dependence as $Lt'$, so that
its contribution to the perturbation expansion is of order
of $O(1/L)$ or $O(1)$. The propagator with momentum
$k\ne 0$ is independent of the momentum, as well as independent of
the parameter $t'$. Therefore, we can put $t'=0$ in the Eq.~(\ref{eq:key}),
and then include the momentum $k=0$ in any of the momentum integral.
The second step only gives rise to an
error of order $1$, so that
\begin{equation}
W_{t'}=W_{t'=0} +O(1),
\end{equation}
where the first term is of the order $O(L)$.
Therefore we see that in the thermodynamic limit,
$W_{t'}$ is independent of the parameter $t'$.

According to
the above argument, the grand thermodynamic potential comes from two
contributions that are completely decoupled. Both parts can be found
in a straightforward way, and we have in the thermodynamic limit
\begin{equation}
\omega = \lim_{L\rightarrow \infty}
{\Omega \over L} = -2t -\beta^{-1}
\ln[2+ z(e^{3\beta J/4} + 3 e^{-\beta J/4}) + 2z^2]
\end{equation}
where $z=e^{\beta \tilde \mu}$ and $\tilde \mu=\mu - t$.
The electron density $n$ of the infinite system, as function
of the chemical potential $\mu$ is given by
$n=-(\partial \omega /\partial \mu)_{\beta}$, i.e.,
\begin{equation}
n=N/L={ [z(e^{3\beta J/4} +3 e^{-\beta J/4}) +4z^2] \over
[2+z(e^{3\beta J/4} + 3 e^{-\beta J/4} ) + 2 z^2 ] }.
\end{equation}
Given electron density $n$, the renormalized
chemical potential $\tilde \mu$ is found from
\begin{eqnarray}
e^{\beta \tilde \mu} ={1\over 2(4-2n) } \{
&&-(1-n) (e^{3\beta J/4} +3 e^{-\beta J/4}) +\nonumber\\
&&+[(1-n)^2 (e^{3\beta J/4} +3 e^{-\beta J/4})^2 + 8 n(4-2n)]^{1/2} \}.
\end{eqnarray}
At half-filling $n=1$, and thus $\tilde \mu=0$.

Let us then calculate the energy density for the infinite
system, i.e., $e=\mu n + {\partial (\beta \omega)
\over \partial \beta }|_{\mu}$.
we obtain
\begin{equation}
e=\mu n  + \{ -2t -{ {z[(3J/4+\tilde \mu)e^{3\beta J/4} + 3 (-J/4+\tilde \mu)
e^{-\beta J/4}] + 4 {\tilde \mu} z^2 }
\over [ 2 + z (e^{3\beta J/4} + 3 e^{-\beta J /4} ) + 2 z^2 ] } \}.
\end{equation}

For fixed number of electron density $n$, taking the low
temperature limit $\beta \rightarrow \infty$, the above becomes the
ground state energy of the system. Let us first consider the
case where the interaction between the local
moment and conduction band is antiferromagnetic, $J>0$.
At less than half-filling $n<1$, we see that
$\tilde
\mu \rightarrow -3J/4 $ when $\beta \rightarrow \infty$. The ground state
energy is thus
\begin{equation}
e_G = -{3\over 4 } J n -2t + tn.
\end{equation}
At half-filling $n=1$, $\tilde \mu=0$, and the ground state energy is
\begin{equation}
e_G =-{3\over 4} J -2t + t.
\end{equation}
At more than half-filling $n>1$, the chemical potential
$\tilde \mu\rightarrow 3J/4$ when
$\beta \rightarrow \infty$, and the ground state energy is given by
\begin{equation}
e_G=-{3\over 4} J n -2t+ t n + {3\over 2} J (n-1).
\end{equation}
{}From these results, we thus obtain
\begin{eqnarray}
&&{d e_G \over d n}|_{n\rightarrow 1^-} = -{3J\over 4} +t\nonumber\\
&&{d e_G \over d n}|_{n\rightarrow 1^+} = {3J\over 4} + t\nonumber\\
&&{d e_G\over d n}|_{n\rightarrow n_0} = -{3J\over 4} + t, n_0<1\nonumber\\
&&{d e_G\over d n}|_{n\rightarrow n_0} = {3J\over 4} + t, n_0>1.
\end{eqnarray}

To study the nature
of the ground state, we employ the idea due to Mattis\cite{mattis1,mattis2},
defining $\mu^+ = E_G(N+1) -E_G(N), \mu^-=
E_G(N)-E_G(N-1)$. The idea is that the system is a metal
if $\delta \mu = \mu^+ - \mu^{-} =0$, while the system
is an insulator if $\delta \mu > 0$.
At half-filling, there is a kink in the chemical potential
$\mu^+> \mu^-$, indicating that the system is an insulator
due to the impurity spins.  At less than half-filling $n< 1$,
$\tilde \mu^+= \tilde \mu^-=-3J/4$, and the system is metallic.
When the filling number is larger than one, $n>1$,
$\tilde \mu^+ = \tilde \mu^-=3J/4$, indicating that the system is metallic.
{}From these analysis, we see that a metal-insulator phase transition
occurs at half-filling, as we turn on the antiferromagnetic interaction
between the local impurity moments and the conduction electrons.
This phase transition would occur even at very small interaction
parameter.

Finally, the analysis of the situation of ferromagnetic coupling
$J<0$ leads to the same conclusion:
At less than half-filling $n<1$,
$\tilde \mu \rightarrow -|J|/4 $ when $\beta \rightarrow \infty$, and
the ground state energy density is
\begin{equation}
e_G= -|J| n/4 + \{-2t\} + tn.
\end{equation}
At exact half-filling, the chemical potential $ \tilde \mu=0$, and the
ground state energy of the system is given by
\begin{equation}
e_G= \{ -2t -|J|/4 \} + t.
\end{equation}
When the electron density is more than 1, we find that
$\tilde \mu \rightarrow |J|/4$ when $\beta \rightarrow \infty$,
and the ground state energy is
\begin{equation}
e_G =|J|n/4 +\{-2t -|J|/2\} + tn.
\end{equation}
{}From these results, we see that the system is metallic if
$n\ne 1$, as indicated by $\delta \mu =\mu^+-\mu^{-} = 0$.
However, at half-filling $n=1$, we have $\delta \mu = |J|/2$,
showing that the system is an insulator.

At zero temperature, when there is no interaction $J=0$, the
system is metallic ( a simple Fermi liquid ). This fixed point
is unstable at half-filling against infinitesimal small interaction.
Our analysis has shown that, at half-filling, a finite charge gap
of the order $|J|$ is open in the excitation spectrum, when turning
on the interaction.
As we have seen, due to the long range aspect of the hopping matrix,
for the thermal potential, the kinetic energy part
decouples with the interaction energy in the thermodynamic limit.
Loosely speaking, at half-filling, to lower the
energy of the system, each conduction electron would
attempt to form a singlet ( triplet ) at each site, when
an antiferromagnetic ( ferromagnetic ) interaction is turned on.
For an electron to hop from a site to any other site on the chain,
even with the long range hopping, it would have to break
two singlet ( triplet )
pairs first, giving rise to a finite charge gap proportional to $|J|$.
In the end, both ferromagnetic and antiferromagnetic interactions will drive
the system to an insulating phase.
In the above study, the diagrammatic
argument does not give us any information on whether one can solve the
Kondo lattice model on a finite size lattice, nor does it provide
explicit wavefunctions of the system. However,
as will be demonstrated below, in the limit
of strong interaction, one can actually find the ground state
wavefunctions rigorously, for a lattice of any size $L$.

In the case $t<0$, at zero temperature,
similar analysis indicates the metal-insulator
phase transition occurs in the system at half-filling, when one
turns on $J$. At finite temperature $T$, the free energy density
is found to be
\begin{equation}
\omega=\lim_{L\rightarrow \infty} {\Omega\over L} = \{ - \beta^{-1} \ln
[ 2 + z (e^{3 \beta J/4} + 3 e^{-\beta J/4} ) + 2 z^2 ]\},
\end{equation}
where $z=e^{\beta (\mu-t)}$ and $\mu$ is the chemical potential.
The diagrammatic analysis has shown that the free energy of the system
can be written in closed form. However, we would like to
emphases that one can not treat the system as if the hopping matrix
were $t=0$, and each electron would form
localized singlet ( triplet) with the impurity at each site for
positive $J$ ( or negative $J$),
uncorrelated with each other.
The long range hopping matrix would delocalize them, and induce
strong correlations between them, as shown below.
Consider the special situation $ t<0, J=+\infty$.
In the limit $J=+\infty$, supposing that there are $N_e$ electrons
on the lattice $L$, with $N_e<L$, each electron will attempt to form
a singlet with the impurity spin at each site, to lower the energy
of the system as much as possible. Some unpaired
impurity spins are left over on the lattice.
Therefore, we work in the Hilbert space where each site
can be either a unpaired impurity spin or a singlet of electron-impurity
bound state. Due to the hopping of the conduction electrons, the singlets
can hop on the lattice.
Let us rewrite the Hamiltonian in the following way:
\begin{equation}
H=|t| \sum_{1\le i\ne j \le L}\sum_{ \sigma=\uparrow,\downarrow}
c_{i\sigma}^\dagger c_{j\sigma} + J \sum_{i=1}^L
c_{i\alpha}^\dagger {\vec \sigma_{\alpha\beta} \over 2} c_{i\beta}
\cdot {\vec S_f(i)}.
\end{equation}
In the limit $J=+\infty$, the basis vectors can be written
as
\begin{eqnarray}
|\alpha>=2^{-N_e/2} \left[ \prod_{i=1}^{N_e} (1-P_{\gamma_i\beta_i}) \right]
&&c_{x_1\gamma_1}^\dagger c_{x_2\gamma_2}^\dagger \cdots
c_{x_{N_e} \gamma_{N_e}}^\dagger |0>\nonumber\\
&&\bigotimes
|\sigma_1, \sigma_2, \cdots, \beta_1, \cdots, \beta_2, \cdots, \sigma_{L-N_e}>,
\end{eqnarray}
where the singlets are located at positions $\{x\}=(x_1<x_2<\cdots<x_{N_e})$,
the unpaired impurity spins $(\sigma_1, \sigma_2, \cdots, \sigma_{L-N_e})$
are positioned at sites $\{y\}=(y_1< y_2<\cdots<y_{L-N_e})$.
Here, the operator $P_{\gamma_i\beta_i}$ permutes the spin indices
$\gamma_i$ and $\beta_i$, to form a singlet of electron and impurity at site
$x_i$.

Denoting the projection operation
onto the $J=+\infty$ subspace by $P$, the projected
Hamiltonian would take the form:
\begin{equation}
\tilde H = P H P = P T P + c = H_1 +c,
\end{equation}
where $T$ is the kinetic energy of the conduction electrons,
the infinite constant $c=(-J/4)N_e$ only shifts the origin of
the energy of the system, a reference energy
which is unimportant physically.
In the space where the z-component of the total spin is fixed,
that is, $S_z=M$, the number of the unpaired up-spin impurities
is $A=M+(L-N_e)/2$, the number of the unpaired down-spin impurities
is $B=-M +(L-N_e)/2$. The dimension of the Hilbert space
associated with $J=+\infty$ is thus
$C_L^{N_e} \times C_{L-N_e}^A$. Any eigenenergy state of the
Hamiltonian $H_1= PTP$ can be written as a linear combination
of the basis vectors,
\begin{equation}
|\phi> =\sum_{\alpha} C(\alpha) |\alpha>.
\end{equation}

To find the ground state wavefunctions,
we may identify the singlets as spinless fermions,
the unpaired impurities as hard core spin 1/2 bosons
hopping on the lattice. Let us consider a system described
by the following Hamiltonian:
\begin{equation}
h=(1/2) \sum_{i\ne j, \sigma} P_G (t_{ij} g_i^\dagger g_j
b_{i\sigma} b_{j\sigma}^\dagger) P_G
\end{equation}
where the $b$ fields are bosonic, $g$ fields are fermionic,
and $t_{ij} = |t|$ for any $i\ne j$. The $b$ fields
commute with the $g$ fields, and $\sum_i g_i^\dagger g_i = N_e$.
The Gutzwiller projector $P_G$ restricts
the system to be in the subspace $g_i^\dagger g_i
+\sum_{\sigma=\uparrow,\downarrow}
b_{i\sigma}^\dagger b_{i\sigma} =1$ for $i=1, 2, \cdots, L$.
For this system, the basis vectors may be represented as follows:
\begin{equation}
|\bar \alpha> = g_{x_1}^\dagger g_{x_2}^\dagger \cdots g_{x_{N_e}}^\dagger
b_{y_1\sigma_1}^\dagger b_{y_2\sigma_2}^\dagger
\cdots b_{y_{L-N_e} \sigma_{L-N_e}}^\dagger |0>.
\end{equation}
One can easily verify the following matrix elements
\begin{equation}
<\beta|H_1|\alpha> = <\bar \beta| h | \bar \alpha>.
\end{equation}
Therefore, the two systems considered here are isomorphic to
each other, and we have the one-to-one correspondence
$|\alpha> \leftrightarrow |\bar \alpha>$ for the basis vectors.
Using the superalgebra representation
\begin{eqnarray}
&&P_F(i) F_{i\sigma}^\dagger P_F(i) =
P_G b_{i\sigma}^\dagger g_i P_G\nonumber\\
&&P_F(i) F_{i\sigma} P_F(i) = P_G b_{i\sigma} g_i^\dagger P_G,
\end{eqnarray}
where
$P_F(i) = (1-F_{i\uparrow}^\dagger F_{i\uparrow}
F_{i\downarrow}^\dagger F_{i\downarrow})$ is the Gutzwiller
operator, and $F$ is a fermionic field,
we have
\begin{equation}
h=(-|t|/2)
\sum_{i\ne j, \sigma} P_F F_{i\sigma}^\dagger F_{j\sigma} P_F,
\end{equation}
where $P_F = \prod_{i=1}^L P_F(i)$, and the number of the $F$ fermions
on the chain is $N_F=L-N_e$.

For the system $ h=(-|t|/2) \sum_{i\ne j, \sigma} P_F
F_{i\sigma}^\dagger F_{j\sigma} P_F$,
it has been proved before
that the ground state energy of the Hamiltonian
with $N_F=L-N_e$ is\cite{verg}
\begin{equation}
e_G = (-|t|) N_e.
\end{equation}
This ground state is highly degenerate, and one has
\begin{equation}
|\phi_G> = P_F F_{0\uparrow}^\dagger F_{0\downarrow}^\dagger
\prod_{i=1, k_i\ne 0 }^{L-N_e -2 } F_{k_i\sigma_i}^\dagger |0>,
\end{equation}
where $F_{0\uparrow}^\dagger, F_{0\downarrow}^\dagger$ and
$F_{k_i\sigma_i}^\dagger$ creat up spin $F$ fermion with momentum
$0$, down-spin $F$ fermion with momentum 0, and $F$ fermion with
spin $\sigma_i$ and momentum $k_i$, respectively.
One can also find some trivial excited energy levels.
For example, one eigenenergy is $ e_1=(|t|/2) (L-N_e) $, another one is
$e_2 = (-|t|/2) N_e$.

We can write out some of the ground state wavefunctions explicitly.
Denote any state vector in the following fashion
\begin{equation}
|\phi> =\sum_{\{X\}, \{Y\}} \Phi (\{X\}, \{Y\} )
\prod_{i=1}^{\bar Q} F_{Y_i\uparrow} \prod_{j=1}^B
F_{X_j\downarrow}^\dagger F_{X_j\uparrow} |P>,
\end{equation}
where the reference state $|P> =
\prod_{i=1}^L F_{i\uparrow}^\dagger |0>$, $\bar Q = N_e$,
$B=-M+(L-N_e)/2$, the amplitude $\Phi$ is antisymmetric
in the positions $\{Y\}$, while symmetric in the positions
$\{X\}$. We have proven that the following Jastrow wavefunctions are
the eigenenergy states with energy $(-|t|) N_e $,
\begin{equation}
\Phi (\{X\}, \{Y\}) = e^{(2\pi i /L) (m_s \sum_{i} X_i + m_h \sum_{j} Y_j )}
\prod_{i<j} d^2(X_i-X_j) \prod_{i<j} d(Y_i-Y_j) \prod_{i, j} d(X_i-Y_j),
\label{eq:wavefunction}
\end{equation}
where the function $d(n)=\sin(\pi n/L)$, and the quantum numbers
$m_s, m_h$ are integers or half-integers, which make sure of the periodic
boundary conditions, and satisfy the following constrains
\begin{eqnarray}
&&(\bar Q +B +1)/2 \le m_h \le L - (\bar Q +B +1)/2\nonumber\\
&& (\bar Q + A +1 )/2 \le (m_h-m_s +L/2)  \le L-(\bar Q + A +1)/2.
\label{eq:condition}
\end{eqnarray}
The effect of the up-spin $F$ fermion hopping operator
can be calculated readily, when it acts on the Jastrow wavefunctions.
Rewriting the Jastrow wavefunctions in terms of
the positions of the up-spin $F$ fermions and the holes,
the resultant amplitude still takes a similar Jastrow product
form. The down-spin $F$ fermion hopping operator effect
can then be handled in the same way.
With these simple calculations,
we have been able to show that
the wavefunctions Eq.~(\ref{eq:wavefunction}) are indeed the eigenfunctions
of the Hamiltonian $h$, with the ground state energy $e_G = (-|t|)N_e$,
as long as the quantum  numbers
$m_s$ and $m_h$ satisfy the constrains Eq.~(\ref{eq:condition}).

It should be remarked that the singlets and the unpaired impurity
spins are strongly correlated with each other, as seen from
the Jastrow product wavefunctions, due to
the long range hopping aspect.
Various correlations between the unpaired impurities
and the singlets can be computed $\it exactly$ in compact form.
Given the Gutzwiller-Jastrow wavefunctions with the
quantum number $m_s$ and $m_h$, we can trivially generalize
Forrester's work $(m_s=L/2, m_h=L/2)$\cite{forrester} to these wavefunctions.
In the conventional case,
where the conduction electrons hop between the nearest
neighboring sites, in the strong interaction limit $J=+\infty$,
the general eigenenergies can be found with the Bethe-ansatz
solution as for the 1D Hubbard model with strong repulsion between
the $F$ fermions.
It was impossible to obtain the closed form of the correlation
functions of the impurity spins and the singlets $\it exactly$,
except their approximate long distance behaviors using
finite size scaling analysis of conformal field theory.
Finally, we would like to note that
these solutions are valid for any lattice size $L$, unlike our first part
of finite $J$ discussion that deals with the large lattice.
Moreover, like in the long range Hubbard model,
the thermodynamic limit and the large $J$ limit
do not commute with each other.

In summary, we have solved exactly the Kondo lattice model in
the thermodynamic limit. Due to
the long range hopping matrix,
we have been able to show that the contribution
of some scattering processes to the thermal potential
scales as $O(1)$ or $O(1/L)$,
while the rest, which can be computed explicitly,
scales as $O(L)$.
The thermodynamics and the ground state energy of the system
have been obtained in this limit.
Our exact solutions demonstrate explicitly the
metal-insulator phase transition at half-filling, as the
interaction of the local impurity moments and conduction band is
switched on at zero temperature.
At any nonzero temperature, the free energy has no singularity,
indicating that there is no phase transition in this system.

We would like to thank the World Laboratory Foundation for
the financial support.
The useful conversations with Prof. P. Coleman,
Dr. B. Doucot and Dr. Q. F. Zhong
are also gratefully acknowledged.
Part of the work was finished
when research was financially supported by the Swiss
National Foundation for Science, to which we feel very
grateful.

\end{document}